\documentclass[aps,pre,twocolumn,groupedaddress,showpacs]{revtex4-1}
\usepackage{amssymb,amsmath,graphicx}

\begin{document}


\title{Relaxation in finite and isolated classical systems: an extension of Onsager's regression hypothesis}

\author{Marcus V. S. Bonan\c{c}a}
\email[]{marcus.bonanca@ufabc.edu.br}
\affiliation{Centro de Ci\^encias Naturais e Humanas, Universidade Federal do ABC, 09210-170, Santo Andr\'e, SP, Brazil}


\date{\today}

\begin{abstract}
In order to derive the reciprocity relations, Onsager formulated a relation between thermal equilibrium fluctuations 
and relaxation widely known as regression hypothesis. It is shown in the present work how such relation can be extended 
to finite and isolated classical systems. This extension is derived from the fluctuation-dissipation theorem for the 
microcanonical ensemble. The results are exemplified with a nonintegrable system in order to motivate possible 
applications to dynamical systems and statistical mechanics of finite systems.
\end{abstract}


\maketitle


\section{Introduction}

In his seminal work on the reciprocity relations\cite{onsager1,*onsager2}, Onsager formulated a hypothesis about the decay 
of thermal equilibrium fluctuations that was essential in his derivation. From this hypothesis, Onsager was able to relate the 
relaxation of nonequilibrium macroscopic observables, obtained from phenomenological equations, to the decay rate of 
equilibrium fluctuations of those quantities. Such a relation between those two apparently different phenomena is however 
quite expected from the point of view of the fluctuation-dissipation theorem derived \cite{callen} 20 years after 
Onsager's work. Indeed, since the derivation of the reciprocity relations from linear response theory, Onsager's 
regression hypothesis is understood as a different but equivalent statement of the fluctuation-dissipation theorem.

On the other hand, Onsager's reciprocity relations have already been extended to far from equilibrium conditions \cite{gallavotti1} 
in different contexts \cite{andrieux} where nonlinear effects are taken into account. Those extensions are based on 
fluctuation theorems which quite often are derived from the so-called \textquotedblleft chaotic hypothesis\textquotedblright 
\cite{gallavotti2,*gallavotti3}. Hence the fluctuation theorems play the role of linear response theory beyond near 
equilibrium conditions and have been also applied to understand the nonequilibrium behavior of systems far from the 
thermodynamic limit \cite{jarzynski}. 

Onsager's original work and its extensions mentioned above always focus on systems in contact with reservoirs. Here 
however we intend to study the near equilibrium behavior of isolated systems when they are finite and standard linear 
response theory cannot be applied. The relaxation to equilibrium of finite and almost isolated quantum systems has been 
studied experimentally using cold atoms \cite{weiss} and has motivated several theoretical works which try to understand 
it and discuss the controversies related to the role played by the nonintegrability in this process 
\cite{rigol,hanggi,gogolin}. For classical systems under such constraints, we show here that a simple extension of the fluctuation-dissipation 
theorem \cite{nielsen,bonanca} shows very clearly the importance of the dynamics in the relaxation 
to equilibrium within a description that is essentially Onsager's regression hypothesis extended to this new situation. An 
important extension of the fluctuation-dissipation theorem for the microcanonical ensemble was first derived by 
Nielsen \cite{nielsen}. It deals with thermodynamic responses to thermal disturbances as for example heat pulses. In 
Ref.\cite{bonanca}, only the response to mechanical disturbances, i.e., those which can be described as additional terms in the 
Hamiltonian, are considered. Since they treat different aspects of the same subject, these works complement each other.

\section{Derivation}

We start presenting the usual regression hypothesis expressed in mathematical terms. Let us consider 
a system described by the following Hamiltonian
\begin{equation}
 H(\lambda_{o} + d\lambda) = H(\lambda_{o}) + d\lambda\,\frac{\partial H}{\partial\lambda}\bigg|_{\lambda = \lambda_{o}}.
\label{eq.1}
\end{equation}
When the value of the parameter $\lambda$ is suddenly switched from $\lambda_{1}=\lambda_{o} + d\lambda$ to 
$\lambda_{o}$ at $t=t_{o}$, the nonequilibrium average value $\bar{B}$ of an observable $B$ of the system evolves 
as \cite{tuckerman}
\begin{equation}
 \bar{B}(t-t_{o}) = \left\langle B \right\rangle_{\lambda_{o}} - 
\frac{d\lambda}{k_{B}T}C_{\lambda_{1}}(t-t_{o}),
\label{eq.2} 
\end{equation}
for $t>t_{o}$, where $\langle\cdot\rangle_{\lambda_{o}}$ denotes the equilibrium average value when 
$\lambda = \lambda_{o}$, $k_{B}$ is the Boltzmann constant, $T$ is the temperature and 
\begin{equation}
 C_{\lambda_{1}}(t) = \langle \delta A(0)\delta B(t)\rangle_{\lambda_{1}}
\label{eq.3}
\end{equation}
is the correlation function with $\delta X(t) = X(t) - \langle X\rangle_{\lambda_{1}}$ and 
$A = (\partial H/\partial\lambda)|_{\lambda=\lambda_{o}}$.

Equation (\ref{eq.2}) is Onsager's regression hypothesis expressed mathematically. It states that the relaxation of $\bar{B}$ to the equilibrium 
value is possible as long as the correlation function $C_{\lambda_{1}}(t)$ decays to zero. On the other hand, if the 
relaxation process can be described by some sort of phenomenological equation, one obtains the decay of 
$C_{\lambda_{1}}(t)$ from (\ref{eq.2}). Since Eq. (\ref{eq.2}) is derived in the usual context of the canonical 
ensemble, the decay of the correlation function is interpreted as a consequence of the heat bath influence. 

We will now derive (\ref{eq.2}) in a different context, namely, when the system is isolated and finite, i.e. there is 
no heat bath and the number of degrees of freedom is such that the system is not in the thermodynamic limit. 
We consider first the system in equilibrium under $H(\lambda_{o}+d\lambda)$. When $t$ is equal to $t_{o}$, $\lambda_{1}$ is 
suddenly switched to $\lambda_{o}$. The system then relaxes to a new equilibrium state. Assuming that the 
system was not far from the final equilibrium state, a linear response calculation describes the relaxation process. 
Although the situation requires the microcanonical ensemble, the fundamental equations of linear response theory do not 
rely on any particular ensemble \cite{kubo}. For $t > t_{o}$ the Hamiltonian is $H(\lambda_{o})$ and the situation can 
be stated as follows: a system, whose Hamiltonian is initially given by (\ref{eq.1}), is in equilibrium when at $t=t_{o}$ 
the generalized force $d\lambda$ is suddenly removed and $H(\lambda_{o}+d\lambda)\to H(\lambda_{o})$. Therefore, from linear 
response theory, one obtains \cite{kubo}
\begin{eqnarray}
 \bar{B}(t-t_{o}) - \langle B\rangle_{\lambda_{1}} 
= d\lambda\int_{-\infty}^{t}ds\,\phi_{BA}(E,t-s)\Theta(s-t_{o}),\nonumber \\
\label{eq.4}
\end{eqnarray}
where $\Theta(x)$ is the step function and $\langle B\rangle_{\lambda_{1}}$ is the following microcanonical average 
over the phase space points $(\mathbf{q},\mathbf{p})$
\begin{equation}
 \langle B\rangle_{\lambda_{1}} = \int d\mathbf{q}\, d\mathbf{p}\, 
\rho_{\lambda_{1}}(\mathbf{q},\mathbf{p})\,B(\mathbf{q},\mathbf{p}),
\label{eq.5}
\end{equation}
with the distribution function $\rho_{\lambda_{1}}$ given by
\begin{eqnarray}
 \rho_{\lambda_{1}}(\mathbf{q},\mathbf{p}) = \frac{\delta\left[E - H(\mathbf{q},\mathbf{p},\lambda_{1})\right]}
{Z_{\lambda_{1}}(E)}, \label{eq.6}
\end{eqnarray}
where $Z_{\lambda_{1}}(E) = \int d\mathbf{q}\, d\mathbf{p}\,\delta\left[E - H(\mathbf{q},\mathbf{p},\lambda_{1})\right]$. 

The response function $\phi_{BA}(E,t-s)$ is \cite{kubo}
\begin{equation}
 \phi_{BA}(E,t-s) = \langle \left\{\delta A(s),\delta B(t)\right\}\rangle_{\lambda_{1}}, 
\label{eq.7}
\end{equation}
where $\left\{\cdot,\cdot\right\}$ is the Poisson bracket, $\delta X(t) = X(\mathbf{q}(t),\mathbf{p}(t)) - 
\langle X\rangle_{\lambda_{1}}$, $(\mathbf{q}(t),\mathbf{p}(t))$ is the solution of Hamilton's 
equations of motion for $N$ degrees of freedom, $B$ is any observable and 
$A = (\partial H/\partial \lambda)|_{\lambda=\lambda_{o}}$.

The fluctuation-dissipation theorem \cite{bonanca} yields then
\begin{equation}
 \tilde{F}_{BA}(z,\omega) = \frac{i}{z\omega}\tilde{\chi}_{BA}(z,\omega),
\label{eq.8}
\end{equation}
where
\begin{eqnarray}
 \lefteqn{\tilde{\chi}_{BA}(z,\omega) = }\nonumber \\
& & \frac{1}{2\pi} \int_{-\infty}^{\infty}d\tau\int_{0}^{\infty}dE\,e^{-(i\omega\tau+Ez)}
    \left[Z_{\lambda}(E)\phi_{BA}(E,\tau)\right],\nonumber\\
\label{eq.9}
\end{eqnarray}
\begin{eqnarray}
 \lefteqn{\tilde{F}_{BA}(z,\omega) = }\nonumber \\
& & \frac{1}{2\pi} \int_{-\infty}^{\infty}d\tau\int_{0}^{\infty}dE\,e^{-(i\omega\tau+Ez)}
    \left[Z_{\lambda}(E)C_{BA}(E,\tau)\right],\nonumber\\
\label{eq.10}
\end{eqnarray}
and $C_{BA}(E,\tau)$, with $\tau = t-s$, is the following correlation function 
\begin{equation}
 C_{BA}(E,t-s) = \langle \delta A(s)\delta B(t)\rangle_{\lambda},
\label{eq.11}
\end{equation}
which differs from (\ref{eq.3}) because of the microcanonical average.

Thus, Eq.(\ref{eq.8}) leads to
\begin{equation}
 \phi_{BA}(E,\tau) = -\frac{1}{Z_{\lambda}(E)}\frac{\partial^{2}}{\partial\tau\partial E}\left[Z_{\lambda}(E)
C_{BA}(E,\tau)\right].
\label{eq.12}
\end{equation}

The integral in (\ref{eq.4}) can be written as
\begin{equation}
 \int_{-\infty}^{t} ds \,\phi_{BA}(t-s)\Theta(s-t_{o})
= \int_{0}^{t-t_{o}}d\tau\, \phi_{BA}(\tau), 
\label{eq.13}
\end{equation}
and from (\ref{eq.12}) and (\ref{eq.13}) one obtains
\begin{eqnarray}
  \int_{0}^{t-t_{o}}d\tau\, \phi_{BA}(\tau)
= -\frac{1}{Z_{\lambda}}\frac{\partial}{\partial E}
\left[Z_{\lambda}C_{BA}(\tau)\right]\bigg|_{0}^{t-t_{o}}
\label{eq.14}
\end{eqnarray}
The dependence on $E$ was omitted for convenience.

Therefore, the relaxation of $B$ to the new equilibrium state is described by the following expression for $t > t_{o}$
\begin{eqnarray}
 \bar{B}(t-t_{o}) = \langle B\rangle_{\lambda_{o}}
-d\lambda\frac{1}{Z_{\lambda_{1}}}\frac{\partial}{\partial E}\left[Z_{\lambda_{1}}C_{BA}(t-t_{o})\right],
\label{eq.15}
\end{eqnarray}
where 
\begin{equation}
 \langle B\rangle_{\lambda_{o}}=\langle B\rangle_{\lambda_{1}}+d\lambda\frac{1}{Z_{\lambda_{1}}}\frac{\partial}
{\partial E}\left[Z_{\lambda_{1}}C_{BA}(0)\right], \label{eq.16}
\end{equation}
since $\lim_{\tau\to\infty}C_{BA}(\tau)=0$ is assumed.

Analogously to Eq. (\ref{eq.2}), the relaxation to equilibrium of $B$ is ruled by a correlation function related to it 
and the relaxation rate is given in terms of the decay rate of equilibrium fluctuations. Thus, the physical contents 
of Eq. (\ref{eq.15}) allow us to interpret it as an extension of Onsager's regression hypothesis to the context of 
finite and isolated classical systems.

Despite the analogy with (\ref{eq.2}), Eq.(\ref{eq.15}) shows that the relaxation rate of $\bar{B}$ [as well as the decay 
rate of $C_{BA}(E,t)$] is given only by the statistical properties of the dynamics produced by $H(\lambda_{o})$. Since the 
system is isolated, there is no influence of external thermal fluctuations on the decay of $C_{BA}(E,t)$ or 
$\bar{B}$ as in (\ref{eq.2}). Therefore, one might ask for which kind of dynamics relaxation occurs. For integrable systems, 
whose motion is quasiperiodic, $C_{BA}(E,t)$ would decay only for $N\to\infty$. Nevertheless, it is well known that 
$C_{BA}(E,t)$ decays for chaotic systems \cite{ruelle1,*ruelle2,gallavotti4,artuso}. For nonintegrable systems, the 
complete spectrum of behaviors, from quasi-periodic to chaotic, could be approximately obtained.

In 1971, van Kampen made severe criticism of linear response theory \cite{kampen} which here, in the context of 
isolated and finite systems, seems to be even harder to answer. However, some of the arguments in the literature 
\cite{kubo,dorfman,evans} supporting standard linear response theory come from dynamical systems theory and are well 
suited for the present discussion. First, it is indeed possible, as mentioned above, that a finite and isolated system 
shows correlation functions which decay with time. In particular, if it has a statistical property called {\it mixing} 
\cite{ruelle1,*ruelle2,dorfman}, it is possible to prove that such decay necessarily happens.
Second, {\it mixing} is also responsible to ensure that an arbitrary smooth distribution in phase space will approach 
the microcanonical one for long times. Thus, although trajectories are very sensitive 
to small perturbations, the time evolution of distributions is rather stable. This would justify the linearization 
procedure leading to Eq. (\ref{eq.4}) at least for a class of systems. There should be also a constraint on time 
scales since the response function (\ref{eq.12}) would be ill defined for times much shorter than the inverse of the decay 
rate of correlations.

\begin{figure}
\centering
\includegraphics[width=0.9\linewidth,clip]{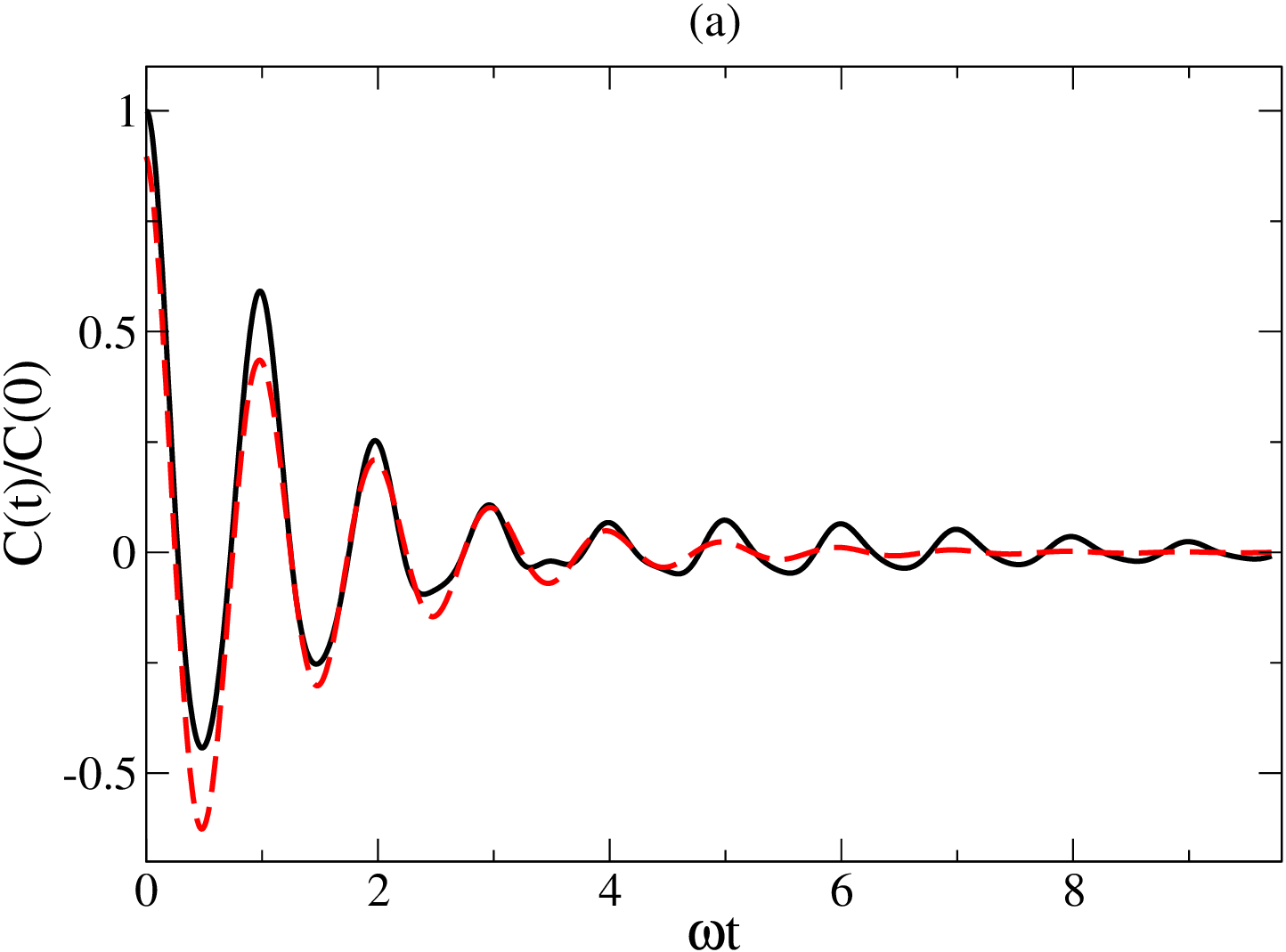}
\includegraphics[width=0.9\linewidth,clip]{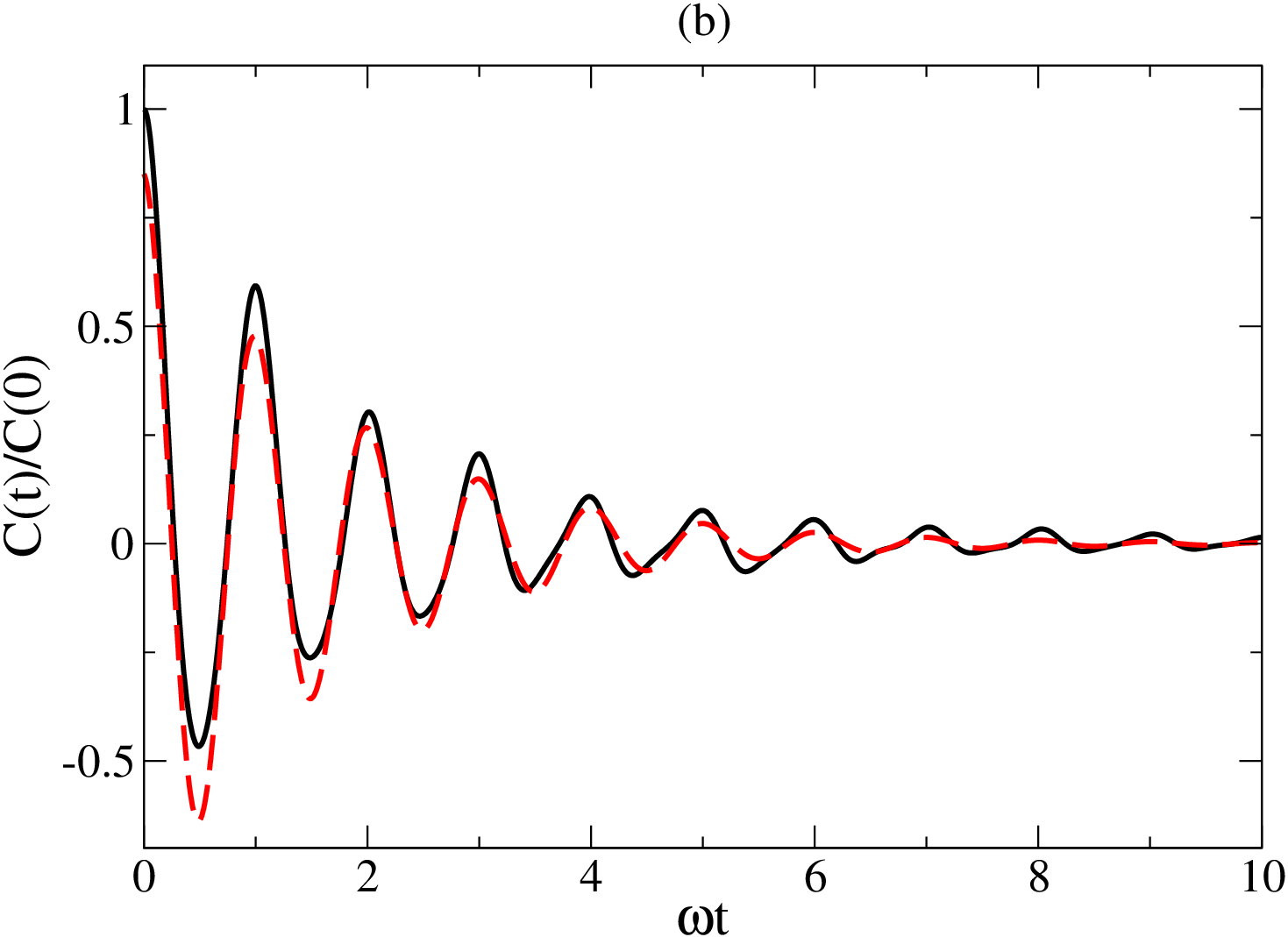}
\caption{Correlation function $C_{BA}(E',t)$ for $B = A = (q_{1}^{4}+q_{2}^{4})/4$ and $E' = 2.5$ 
The solid black lines were obtained numerically for $4\times10^{5}$ initial conditions. The dashed red lines are the 
fitting of $\mathcal{A}'e^{-\alpha't}\cos{(\omega't)}$. (a) $\lambda_{1}=0.1$, $\mathcal{A}'=21.5$, $\alpha' = 0.20$ 
and $\omega' = 1.7$. (b) $\lambda_{1}=0.12$, $\mathcal{A}'=14.5$, $\alpha' = 0.17$ and $\omega' = 1.8$.}\label{fig1}
\end{figure}
\begin{figure}
\centering
\includegraphics[width=0.9\linewidth,clip]{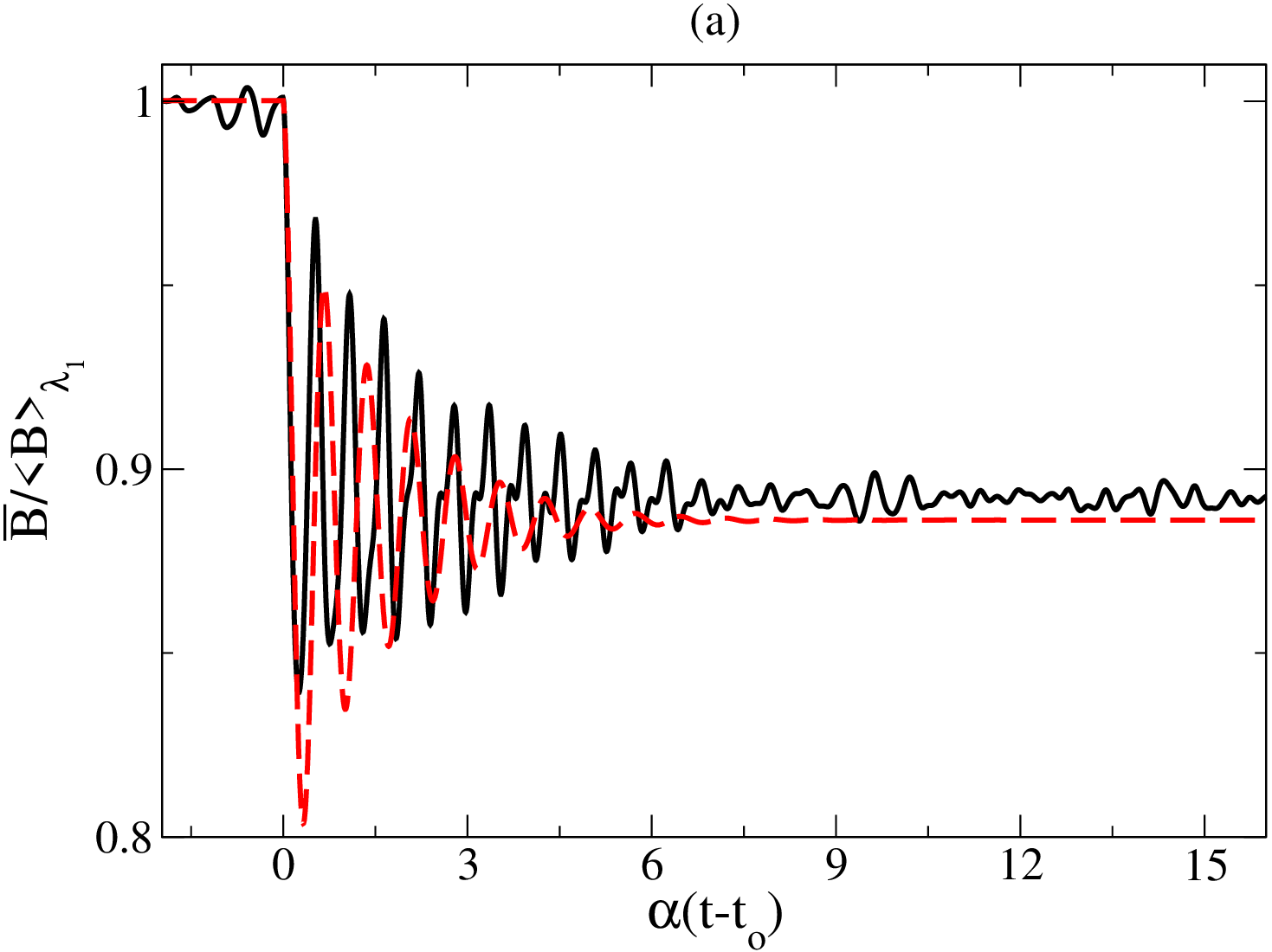}
\includegraphics[width=0.9\linewidth,clip]{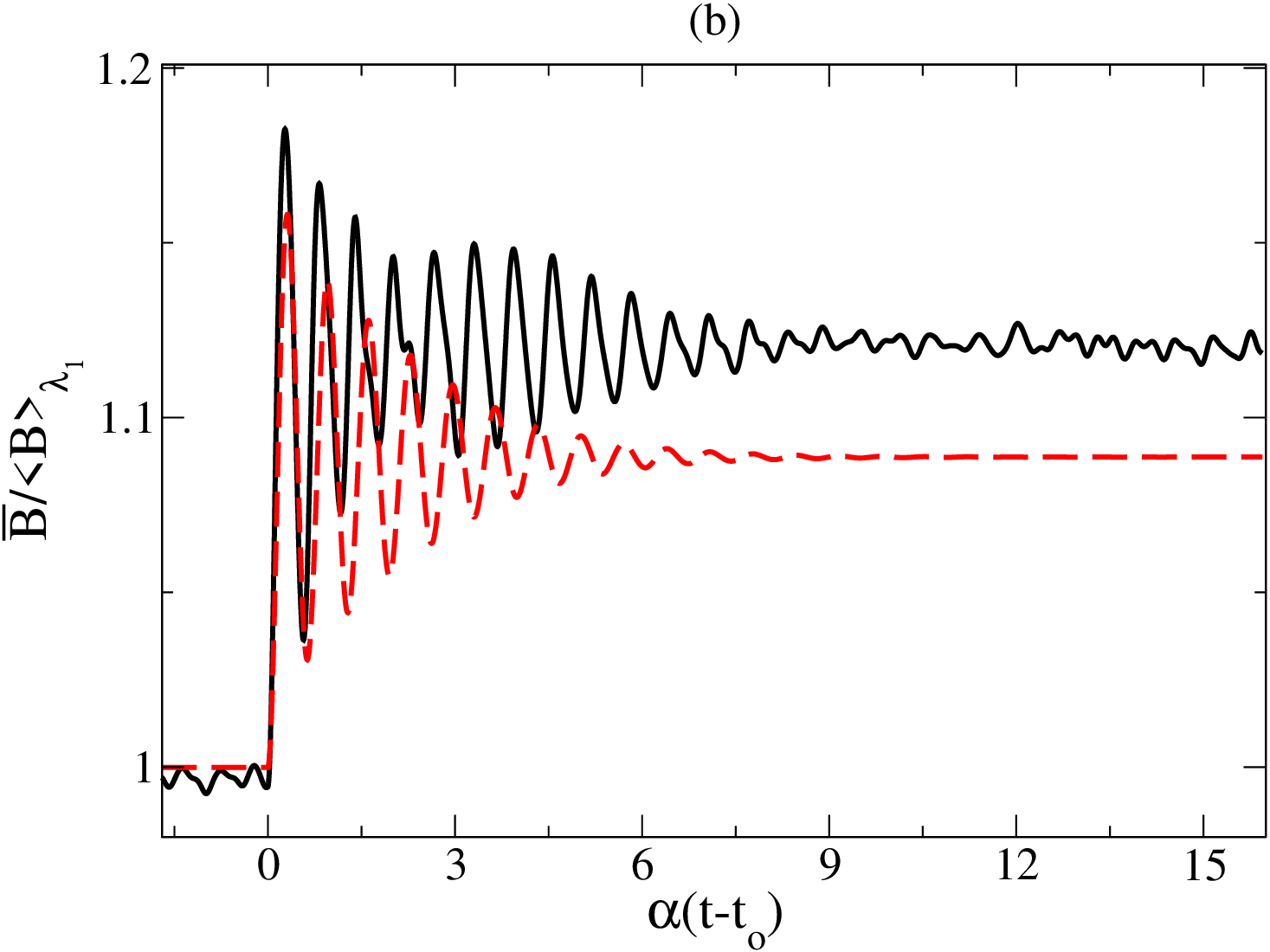}
\caption{Relaxation of $\bar{B}$ for $B = (q_{1}^{4}+q_{2}^{4})/4$ and initial energy $E = 5.0$. The solid black lines 
were obtained numerically for $4\times10^{5}$ initial conditions. The dashed red lines are the predictions of Eq. (\ref{eq.15}) 
using $C_{BA}$ obtained numerically and the scaling property. (a) $\lambda_{1}=0.1$ switched to $\lambda_{o}=0.12$
and the value of $\langle B\rangle_{\lambda_{1}}=7.54$ obtained numerically for $4\times10^{5}$ initial conditions
with $E=5.0$ and $\lambda_{1}=0.1$. (b) $\lambda_{1}=0.12$ switched to $\lambda_{o} = 0.1$ and the value of 
$\langle B\rangle_{\lambda_{1}}=6.55$ obtained numerically for $4\times10^{5}$ initial conditions with $E=5.0$ and 
$\lambda_{1}=0.12$}\label{fig2}
\end{figure}
%

\section{Example}

In order to motivate possible applications of Eq. (\ref{eq.15}) to both low-dimensional dynamical systems and statistical 
mechanics of finite systems, we will consider the relaxation process in a system described by the following Hamiltonian \cite{percival}
\begin{equation}
 H = \frac{p_{1}^{2}}{2}+\frac{p_{2}^{2}}{2}+\frac{q_{1}^{2}q_{2}^{2}}{2}+\lambda\frac{(q_{1}^{4}+q_{2}^{4})}{4},
\label{eq.17}
\end{equation}
which is integrable only for $\lambda = 1$. It can be verified through the Poincar\'e surface of sections that the motion 
generated by (\ref{eq.17}) gets less and less regular as $\lambda$ decreases from the value of unity. If on one hand 
the relaxation process is very well defined for low-dimensional hyperbolic systems, on the other hand most of the realistic models used 
in statistical mechanics of classical systems are nonintegrable. For a nonintegrable systems with as few degrees of freedom 
as (\ref{eq.17}), relaxation may or may not happen and one has to find numerically the range of parameters where it occurs.
In our application of (\ref{eq.15}) to this case, we have chosen $A=(\partial H/\partial \lambda)|_{\lambda=\lambda_{o}} =
(q_{1}^{4}+q_{2}^{4})/4$ and $B=A$. There is a small range of values of $\lambda$ ($0.9\lesssim\lambda\lesssim 0.13$) where 
$C_{BA}(\tau)$ can be fitted by the expression $\mathcal{A}e^{-\alpha\tau}\cos{(\omega\tau)}$ (see Fig.\ref{fig1}). It is 
not our aim here to find the exact functional form of the correlation function. Instead we want to find out whether a 
simple description of it (as the one just written above) is enough to describe the relaxation process approximately.

The dynamics given by (\ref{eq.17}) is scalable with energy, i.e. the Hamilton equations remain invariant under a 
transformation $(q'_{1,2},p'_{1,2},t')\to(q_{1,2},p_{1,2},t)$ given by the equations
\begin{equation}
 \frac{q_{1,2}}{q'_{1,2}} = \left(\frac{E}{E'}\right)^{1/4},\; 
\frac{p_{1,2}}{p'_{1,2}} = \left(\frac{E}{E'}\right)^{1/4},\;
\frac{t}{t'} = \left(\frac{E'}{E}\right)^{1/4},
\end{equation}
This property of (\ref{eq.17}) yields $\mathcal{A}/\mathcal{A}'=(E/E')^{2}$ since 
$(q_{1}^{4}+q_{2}^{4})^{2}\propto(E/E')^{2}$, $\alpha/\alpha'=(E/E')^{1/4}$ and 
$\omega/\omega'=(E/E')^{1/4}$. One can also verify that $Z_{\lambda}(E)\propto E^{1/2}$.
Therefore, the scaling with energy allows us to perform the derivative in (\ref{eq.15}) taking a certain value $E'$ 
as the reference and obtaining $C_{BA}(E,t)$ from the parameters of $C_{BA}(E',t)$, keeping of course the same functional 
form and the same value of $\lambda$. In summary, the system described by (\ref{eq.17}) was chosen as an example 
because it allows a simple illustration of (\ref{eq.15}) applied to extremely few degrees of freedom. For models with 
Lennard-Jones potentials, for example, the dependence of correlation functions with energy is much more complicated 
and is accessible only numerically. Besides, it has been shown recently \cite{marchiori} that a finite 
collection of (\ref{eq.17}) can act as an environment that induces relaxation on a simple degree of freedom.

\section{Discussion and Conclusions}

All the numerical results were obtained from the integration of the equations of motion of (\ref{eq.17}) using a 
fourth-order sympletic integrator \cite{forest}. In Fig.\ref{fig1} is shown the numerical results for 
$C_{BA}(E',t)$. Although it is clear that the fitting is not excellent, one obtains afterwards a good agreement 
between numerical and analytical results for the relaxation of $\bar{B}$. 
The analytical results should indeed be called semi-analytical since both $\langle B\rangle_{\lambda_{1}}$ and $C_{BA}$ 
were obtained numerically. In Fig.\ref{fig2}, $\bar{B}(t-t_{o})$ is the result of the average over several $B(t-t_{o})$ 
obtained from the time evolution of initial conditions distributed over an energy surface with $E=5.0$. In Fig.\ref{fig2}(a) 
a sudden switching of $\lambda_{1}=0.1$ to $\lambda_{o}=0.12$ at $t_{o}$ leads to the relaxation of $\bar{B}$ to a new 
equilibrium value. Although amplitude and frequency of oscillations are not correctly described, the relaxation time and 
the value of $\langle B\rangle_{\lambda_{o}}$ are well predicted. In Fig.\ref{fig2}(b), the comparison between numerical 
and analytical results for $\bar{B}$ is shown for $\lambda_{1}=0.12$ switched to $\lambda_{o}=0.1$. As before, amplitude and 
frequency of oscillations are roughly described and the value predicted for $\langle B\rangle_{\lambda_{o}}$ is not 
as good as in Fig.\ref{fig2}(a). The relaxation time however is still in good agreement with the numerical results.

In conclusion, we have derived an extension of Onsager's regression hypothesis from linear response theory when the
system of interest is isolated and finite. Although thermal fluctuations induced by a heat bath are absent in this context, 
the expression obtained relates, as usual, relaxation to equilibrium fluctuations. Hence the new feature is that the 
decay of correlations is given only by the instrinsic dynamics of the system. The relaxation of a nonintegrable systems
with two degress of freedom has illustrated that. The outlook is to extend the present result to the quantum 
regime.

\begin{acknowledgments}
The author acknowledges support of UFABC. The author is also grateful to M. de Koning and R. Venegeroles for 
valuable discussions and suggestions. 
\end{acknowledgments}


\begin{thebibliography}{10}%
\makeatletter
\providecommand \@ifxundefined [1]{%
 \ifx #1\undefined \expandafter \@firstoftwo
 \else \expandafter \@secondoftwo
\fi
}%
\providecommand \@ifnum [1]{%
 \ifnum #1\expandafter \@firstoftwo
 \else \expandafter \@secondoftwo
\fi
}%
\providecommand \enquote [1]{``#1''}%
\providecommand \bibnamefont  [1]{#1}%
\providecommand \bibfnamefont [1]{#1}%
\providecommand \citenamefont [1]{#1}%
\providecommand\href[0]{\@sanitize\@href}%
\providecommand\@href[1]{\endgroup\@@startlink{#1}\endgroup\@@href}%
\providecommand\@@href[1]{#1\@@endlink}%
\providecommand \@sanitize [0]{\begingroup\catcode`\&12\catcode`\#12\relax}%
\@ifxundefined \pdfoutput {\@firstoftwo}{%
 \@ifnum{\z@=\pdfoutput}{\@firstoftwo}{\@secondoftwo}%
}{%
 \providecommand\@@startlink[1]{\leavevmode}%
 \providecommand\@@endlink[0]{}%
}{%
 \providecommand\@@startlink[1]{%
  \leavevmode
  \pdfstartlink
   attr{/Border[0 0 1 ]/H/I/C[0 1 1]}%
   user{/Subtype/Link/A<</Type/Action/S/URI/URI(#1)>>}%
  \relax
 }%
 \providecommand\@@endlink[0]{\pdfendlink}%
}%
\providecommand \url  [0]{\begingroup\@sanitize \@url }%
\providecommand \@url [1]{\endgroup\@href {#1}{\urlprefix}}%
\providecommand \urlprefix [0]{URL }%
\providecommand \Eprint[0]{\href }%
\@ifxundefined \urlstyle {%
  \providecommand \doi [1]{doi:\discretionary{}{}{}#1}%
}{%
  \providecommand \doi [0]{doi:\discretionary{}{}{}\begingroup
  \urlstyle{rm}\Url }%
}%
\providecommand \doibase [0]{http://dx.doi.org/}%
\providecommand \Doi[1]{\href{\doibase#1}}%
\providecommand \bibAnnote [3]{%
  \BibitemShut{#1}%
  \begin{quotation}\noindent
    \textsc{Key:}\ #2\\\textsc{Annotation:}\ #3%
  \end{quotation}%
}%
\providecommand \bibAnnoteFile [2]{%
  \IfFileExists{#2}{\bibAnnote {#1} {#2} {\input{#2}}}{}%
}%
\providecommand \typeout [0]{\immediate \write \m@ne }%
\providecommand \selectlanguage [0]{\@gobble}%
\providecommand \bibinfo [0]{\@secondoftwo}%
\providecommand \bibfield [0]{\@secondoftwo}%
\providecommand \translation [1]{[#1]}%
\providecommand \BibitemOpen[0]{}%
\providecommand \bibitemStop [0]{}%
\providecommand \bibitemNoStop [0]{.\EOS\space}%
\providecommand \EOS [0]{\spacefactor3000\relax}%
\providecommand \BibitemShut [1]{\csname bibitem#1\endcsname}%
\bibitem{onsager1}%
  \BibitemOpen
  \bibfield{author}{%
  \bibinfo {author} {\bibfnamefont{L.}~\bibnamefont{Onsager}},\ }%
  \bibfield{journal}{%
  \bibinfo {journal} {Phys. Rev.}\ }%
  \textbf{\bibinfo {volume} {37}},\ \bibinfo {pages} {405} (\bibinfo {year}
  {1931})%
  \bibAnnoteFile{NoStop}{onsager1}%
\bibitem{onsager2}%
  \BibitemOpen
  \bibfield{author}{%
  \bibinfo {author} {\bibfnamefont{L.}~\bibnamefont{Onsager}},\ }%
  \bibfield{journal}{%
  \bibinfo {journal} {Phys. Rev.}\ }%
  \textbf{\bibinfo {volume} {38}},\ \bibinfo {pages} {2265} (\bibinfo {year}
  {1931})%
  \bibAnnoteFile{NoStop}{onsager2}%
\bibitem{callen}%
  \BibitemOpen
  \bibfield{author}{%
  \bibinfo {author} {\bibfnamefont{H.~B.}\ \bibnamefont{Callen}}\ and\ \bibinfo
  {author} {\bibfnamefont{T.~A.}\ \bibnamefont{Welton}},\ }%
  \bibfield{journal}{%
  \bibinfo {journal} {Phys. Rev.}\ }%
  \textbf{\bibinfo {volume} {83}},\ \bibinfo {pages} {34} (\bibinfo {year}
  {1951})%
  \bibAnnoteFile{NoStop}{callen}%
\bibitem{gallavotti1}%
  \BibitemOpen
  \bibfield{author}{%
  \bibinfo {author} {\bibfnamefont{G.}~\bibnamefont{Gallavotti}},\ }%
  \bibfield{journal}{%
  \bibinfo {journal} {Phys. Rev. Lett.}\ }%
  \textbf{\bibinfo {volume} {77}},\ \bibinfo {pages} {4334} (\bibinfo {year}
  {1996})%
  \bibAnnoteFile{NoStop}{gallavotti1}%
\bibitem{andrieux}%
  \BibitemOpen
  \bibfield{author}{%
  \bibinfo {author} {\bibfnamefont{D.}~\bibnamefont{Andrieux}}\ and\ \bibinfo
  {author} {\bibfnamefont{P.}~\bibnamefont{Gaspard}},\ }%
  \bibfield{journal}{%
  \bibinfo {journal} {J. Stat. Mech.}\ }%
  \textbf{\bibinfo {volume} {2007}},\ \bibinfo {pages} {P02006} (\bibinfo
  {year} {2007})%
  \bibAnnoteFile{NoStop}{andrieux}%
\bibitem{gallavotti2}%
  \BibitemOpen
  \bibfield{author}{%
  \bibinfo {author} {\bibfnamefont{G.}~\bibnamefont{Gallavotti}}\ and\ \bibinfo
  {author} {\bibfnamefont{E.~G.~D.}\ \bibnamefont{Cohen}},\ }%
  \bibfield{journal}{%
  \bibinfo {journal} {Phys. Rev. Lett.}\ }%
  \textbf{\bibinfo {volume} {74}},\ \bibinfo {pages} {2694} (\bibinfo {year}
  {1995})%
  \bibAnnoteFile{NoStop}{gallavotti2}%
\bibitem{gallavotti3}%
  \BibitemOpen
  \bibfield{author}{%
  \bibinfo {author} {\bibfnamefont{G.}~\bibnamefont{Gallavotti}}\ and\ \bibinfo
  {author} {\bibfnamefont{E.~G.~D.}\ \bibnamefont{Cohen}},\ }%
  \bibfield{journal}{%
  \bibinfo {journal} {J. Stat. Phys.}\ }%
  \textbf{\bibinfo {volume} {80}},\ \bibinfo {pages} {931} (\bibinfo {year}
  {1995})%
  \bibAnnoteFile{NoStop}{gallavotti3}%
\bibitem{jarzynski}%
  \BibitemOpen
  \bibfield{author}{%
  \bibinfo {author} {\bibfnamefont{D.}~\bibnamefont{Collin}}, \bibinfo {author}
  {\bibfnamefont{F.}~\bibnamefont{Ritort}}, \bibinfo {author}
  {\bibfnamefont{C.}~\bibnamefont{Jarzynski}}, \bibinfo {author}
  {\bibfnamefont{S.~B.}\ \bibnamefont{Smith}}, \bibinfo {author}
  {\bibfnamefont{J.}~\bibnamefont{I.~Tinoco}},\ and\ \bibinfo {author}
  {\bibfnamefont{C.}~\bibnamefont{Bustamante}},\ }%
  \bibfield{journal}{%
  \bibinfo {journal} {Nature}\ }%
  \textbf{\bibinfo {volume} {437}},\ \bibinfo {pages} {231} (\bibinfo {year}
  {2005})%
  \bibAnnoteFile{NoStop}{jarzynski}%
\bibitem{weiss}%
  \BibitemOpen
  \bibfield{author}{%
  \bibinfo {author} {\bibfnamefont{T.}~\bibnamefont{Kinoshita}}, \bibinfo
  {author} {\bibfnamefont{T.}~\bibnamefont{Wenger}},\ and\ \bibinfo {author}
  {\bibfnamefont{D.~S.}\ \bibnamefont{Weiss}},\ }%
  \bibfield{journal}{%
  \bibinfo {journal} {Nature}\ }%
  \textbf{\bibinfo {volume} {440}},\ \bibinfo {pages} {900} (\bibinfo {year}
  {2006})%
  \bibAnnoteFile{NoStop}{weiss}%
\bibitem{rigol}%
  \BibitemOpen
  \bibfield{author}{%
  \bibinfo {author} {\bibfnamefont{M.}~\bibnamefont{Rigol}}, \bibinfo {author}
  {\bibfnamefont{V.}~\bibnamefont{Dunjko}},\ and\ \bibinfo {author}
  {\bibfnamefont{M.}~\bibnamefont{Olshanii}},\ }%
  \bibfield{journal}{%
  \bibinfo {journal} {Nature}\ }%
  \textbf{\bibinfo {volume} {452}},\ \bibinfo {pages} {854} (\bibinfo {year}
  {2008})%
  \bibAnnoteFile{NoStop}{rigol}%
\bibitem{hanggi}%
  \BibitemOpen
  \bibfield{author}{%
  \bibinfo {author} {\bibfnamefont{A.~V.}\ \bibnamefont{Ponomarev}}, \bibinfo
  {author} {\bibfnamefont{S.}~\bibnamefont{Denisov}},\ and\ \bibinfo {author}
  {\bibfnamefont{P.}~\bibnamefont{H{\"{a}}nggi}},\ }%
  \bibfield{journal}{%
  \bibinfo {journal} {Phys. Rev. Lett.}\ }%
  \textbf{\bibinfo {volume} {106}},\ \bibinfo {pages} {010405} (\bibinfo {year}
  {2011})%
  \bibAnnoteFile{NoStop}{hanggi}%
\bibitem{gogolin}%
  \BibitemOpen
  \bibfield{author}{%
  \bibinfo {author} {\bibfnamefont{C.}~\bibnamefont{Gogolin}}, \bibinfo
  {author} {\bibfnamefont{M.~P.}\ \bibnamefont{M{\"{u}}ller}},\ and\ \bibinfo
  {author} {\bibfnamefont{J.}~\bibnamefont{Eisert}},\ }%
  \bibfield{journal}{%
  \bibinfo {journal} {Phys. Rev. Lett.}\ }%
  \textbf{\bibinfo {volume} {106}},\ \bibinfo {pages} {040401} (\bibinfo {year}
  {2011})%
  \bibAnnoteFile{NoStop}{gogolin}%
\bibitem{nielsen}%
  \BibitemOpen
  \bibfield{author}{%
  \bibinfo {author} {\bibfnamefont{J.~K.}\ \bibnamefont{Nielsen}},\ }%
  \bibfield{journal}{%
  \bibinfo {journal} {Phys. Rev. E}\ }%
  \textbf{\bibinfo {volume} {60}},\ \bibinfo {pages} {471} (\bibinfo {year}
  {1999})%
  \bibAnnoteFile{NoStop}{nielsen}%
\bibitem{bonanca}%
  \BibitemOpen
  \bibfield{author}{%
  \bibinfo {author} {\bibfnamefont{M.~V.~S.}\ \bibnamefont{Bonan{\c{c}}a}},\ }%
  \bibfield{journal}{%
  \bibinfo {journal} {Phys. Rev. E}\ }%
  \textbf{\bibinfo {volume} {78}},\ \bibinfo {pages} {031107} (\bibinfo {year}
  {2008})%
  \bibAnnoteFile{NoStop}{bonanca}%
\bibitem{tuckerman}%
  \BibitemOpen
  \bibfield{author}{%
  \bibinfo {author} {\bibfnamefont{M.~E.}\ \bibnamefont{Tuckerman}},\ }%
  \emph{\bibinfo {title} {Statistical Mechanics: Theory and Molecular
  Simulation}}\ (\bibinfo {publisher} {Oxford University Press},\ \bibinfo
  {address} {New York},\ \bibinfo {year} {2010})%
  \bibAnnoteFile{NoStop}{tuckerman}%
\bibitem{kubo}%
  \BibitemOpen
  \bibfield{author}{%
  \bibinfo {author} {\bibfnamefont{R.}~\bibnamefont{Kubo}}, \bibinfo {author}
  {\bibfnamefont{M.}~\bibnamefont{Toda}},\ and\ \bibinfo {author}
  {\bibfnamefont{N.}~\bibnamefont{Hashitsume}},\ }%
  \emph{\bibinfo {title} {Statistical Physics II}}\ (\bibinfo {publisher}
  {Springer-Verlag},\ \bibinfo {address} {Berlin},\ \bibinfo {year} {1985})%
  \bibAnnoteFile{NoStop}{kubo}%
\bibitem{ruelle1}%
  \BibitemOpen
  \bibfield{author}{%
  \bibinfo {author} {\bibfnamefont{D.}~\bibnamefont{Ruelle}},\ }%
  \bibfield{journal}{%
  \bibinfo {journal} {Phys. Rev. Lett.}\ }%
  \textbf{\bibinfo {volume} {56}},\ \bibinfo {pages} {405} (\bibinfo {year}
  {1986})%
  \bibAnnoteFile{NoStop}{ruelle1}%
\bibitem{ruelle2}%
  \BibitemOpen
  \bibfield{author}{%
  \bibinfo {author} {\bibfnamefont{D.}~\bibnamefont{Ruelle}},\ }%
  \bibfield{journal}{%
  \bibinfo {journal} {J. Stat. Phys.}\ }%
  \textbf{\bibinfo {volume} {44}},\ \bibinfo {pages} {281} (\bibinfo {year}
  {1986})%
  \bibAnnoteFile{NoStop}{ruelle2}%
\bibitem{gallavotti4}%
  \BibitemOpen
  \bibfield{author}{%
  \bibinfo {author} {\bibfnamefont{P.~L.}\ \bibnamefont{Garrido}}\ and\
  \bibinfo {author} {\bibfnamefont{G.}~\bibnamefont{Gallavotti}},\ }%
  \bibfield{journal}{%
  \bibinfo {journal} {J. Stat. Phys.}\ }%
  \textbf{\bibinfo {volume} {76}},\ \bibinfo {pages} {549} (\bibinfo {year}
  {1994})%
  \bibAnnoteFile{NoStop}{gallavotti4}%
\bibitem{artuso}%
  \BibitemOpen
  \bibfield{author}{%
  \bibinfo {author} {\bibfnamefont{R.}~\bibnamefont{Artuso}}, \bibinfo {author}
  {\bibfnamefont{G.}~\bibnamefont{Casati}},\ and\ \bibinfo {author}
  {\bibfnamefont{I.}~\bibnamefont{Guarneri}},\ }%
  \bibfield{journal}{%
  \bibinfo {journal} {J. Stat. Phys.}\ }%
  \textbf{\bibinfo {volume} {83}},\ \bibinfo {pages} {145} (\bibinfo {year}
  {1996})%
  \bibAnnoteFile{NoStop}{artuso}%
\bibitem{kampen}%
  \BibitemOpen
  \bibfield{author}{%
  \bibinfo {author} {\bibfnamefont{N.~G.}\ \bibnamefont{van Kampen}},\ }%
  \bibfield{journal}{%
  \bibinfo {journal} {Phys. Norv.}\ }%
  \textbf{\bibinfo {volume} {5}},\ \bibinfo {pages} {279} (\bibinfo {year}
  {1971})%
  \bibAnnoteFile{NoStop}{kampen}%
\bibitem{dorfman}%
  \BibitemOpen
  \bibfield{author}{%
  \bibinfo {author} {\bibfnamefont{J.~R.}\ \bibnamefont{Dorfman}},\ }%
  \emph{\bibinfo {title} {An Introduction to Chaos in Nonequilibrium
  Statistical Mechanics}}\ (\bibinfo {publisher} {Cambridge University Press},\
  \bibinfo {address} {Cambridge},\ \bibinfo {year} {1999})%
  \bibAnnoteFile{NoStop}{dorfman}%
\bibitem{evans}%
  \BibitemOpen
  \bibfield{author}{%
  \bibinfo {author} {\bibfnamefont{D.~J.}\ \bibnamefont{Evans}}\ and\ \bibinfo
  {author} {\bibfnamefont{G.}~\bibnamefont{Morriss}},\ }%
  \emph{\bibinfo {title} {Statistical Mechanics of Nonequilibrium Liquids}}\
  (\bibinfo {publisher} {Cambridge University Press},\ \bibinfo {address}
  {Cambridge},\ \bibinfo {year} {2008})%
  \bibAnnoteFile{NoStop}{evans}%
\bibitem{percival}%
  \BibitemOpen
  \bibfield{author}{%
  \bibinfo {author} {\bibfnamefont{A.}~\bibnamefont{Carnegie}}\ and\ \bibinfo
  {author} {\bibfnamefont{I.~C.}\ \bibnamefont{Percival}},\ }%
  \bibfield{journal}{%
  \bibinfo {journal} {J. Phys. A: Math. Gen.}\ }%
  \textbf{\bibinfo {volume} {17}},\ \bibinfo {pages} {801} (\bibinfo {year}
  {1984})%
  \bibAnnoteFile{NoStop}{percival}%
\bibitem{marchiori}%
  \BibitemOpen
  \bibfield{author}{%
  \bibinfo {author} {\bibfnamefont{M.~A.}\ \bibnamefont{Marchiori}}\ and\
  \bibinfo {author} {\bibfnamefont{M.~A.~M.}\ \bibnamefont{de~Aguiar}},\ }%
  \bibfield{journal}{%
  \bibinfo {journal} {Phys. Rev. E}\ }%
  \textbf{\bibinfo {volume} {83}},\ \bibinfo {pages} {061112} (\bibinfo {year}
  {2011})%
  \bibAnnoteFile{NoStop}{marchiori}%
\bibitem{forest}%
  \BibitemOpen
  \bibfield{author}{%
  \bibinfo {author} {\bibfnamefont{E.}~\bibnamefont{Forest}}\ and\ \bibinfo
  {author} {\bibfnamefont{R.~D.}\ \bibnamefont{Ruth}},\ }%
  \bibfield{journal}{%
  \bibinfo {journal} {Physica D}\ }%
  \textbf{\bibinfo {volume} {43}},\ \bibinfo {pages} {105} (\bibinfo {year}
  {1990})%
  \bibAnnoteFile{NoStop}{forest}%
\end{thebibliography}

%

\end{document}